\documentclass[a4paper,10pt,fleqn]{article}

\usepackage[english]{babel}
\usepackage{inputenc}
\usepackage{amsmath,amsfonts,amssymb,marvosym}
\usepackage[colorlinks=true,linkcolor=blue,citecolor=blue,urlcolor=blue]{hyperref}
\usepackage{breakurl}
\usepackage{graphicx}
\usepackage{tabularx}
\usepackage{lscape}
\usepackage{cite}
\usepackage{times}
\usepackage{pifont}
\usepackage{enumerate}
\usepackage[]{lipsum}
\usepackage[mathlines]{lineno}
\usepackage{blindtext}
\usepackage{color}
\usepackage{blindtext}
\usepackage{wrapfig}
\usepackage[textwidth=17cm,textheight=23cm]{geometry}

\definecolor{newred}{RGB}{221,24,31}
\definecolor{newblue}{RGB}{0,96,173}
\definecolor{newgreen}{RGB}{0,128,0}
\definecolor{newgrey}{RGB}{180,180,180}

\usepackage{draftwatermark}
\SetWatermarkText{\Huge arXiv version - \today}

\title{Effect of an advanced model-based iterative reconstruction algorithm on texture and visual impression of images privieded by a dual source CT scanner}

\author{Babak Alikhani PhD\footnote{Corresponding author:babak.alikhani$@$diakovere.de}~ $^{1,2}$~, Thomas Werncke MD$^{1}$~, Hans-J\"urgen Raatschen MD$^{1}$~,\\ Frank Wacker MD$^{1}$~, Hoen-oh Shin MD$^{1}$ \\{\small $^{1}$ Institute for Diagnostic and Interventional Radiology, Hannover Medical School, Hannover, Germany} \\ {\small $^{2}$ Center for Radiology and Nuclear Medicine, DIAKOVERE gGmbH, Hannover, Germany}}

\date{\today}

\begin{document}
\maketitle

\begin{abstract}\noindent
\centering
\begin{minipage}{0.7\textwidth}
\textbf{Purpose}: To evaluate the influence of a model-based iterative reconstruction algorithm (ADMIRE) on image texture and image visual impression as a supplement to measurements of common image quality parameters such as noise levels and spatial resolution. \\ \\
\textbf{Methods}: An American College of Radiology CT accreditation phantom (ACR phantom) was examined at different radiation dose levels expressed by the volumetric computed tomography dose index ($\mathrm{CTDI_{vol}}=$ 0.67, 1.64, 3.31 and 6.65~mGy). To characterize the image texture, two Haralick texture parameters, i.e. contrast and entropy, for different dose level and reconstruction algorithms were assessed. The visual impression of images and their structural differences were evaluated using the structural similarity index (SSIM). Noise defined as the standard deviation of the voxel density was determined for all dose and AMIRE levels and compared to those by filtered back projection. The spatial resolution was determined by the modulation transfer functions and the line spread function. \\ \\
\textbf{Results}: The Haralick texture parameters, contrast and entropy, decreased with increasing ADMIRE levels I up to V. ADMIRE III, IV and V offered a comparable contrast and entropy to those calculated by the filter back projection with a radiation dose reduction up to 50\%. On the other hand, $\mathrm{SSIM}$ improved with increasing ADMIRE levels.
$\mathrm{SSIM}$ calculated by ADMIRE IV and V showed similar values by the filter back projection with a $\mathrm{CTDI_{vol}}$ decrease up to 50\%. Spatial resolution was retained up to 90\% dose reduction. With increasing ADMIRE as well as dose level the noise distribution shifted to a more narrow distribution, which was in accordance with the reconstructed images using ADMIRE. \\ \\
\textbf{Conclusion}: Texture analysis and $\mathrm{SSIM}$ allow a more realistic assessment of the dose reduction potential of iterative reconstruction algorithms than quality metrics only based on physical measurements of noise distribution or spatial resolution. This work presented that by means of the ADMIRE algorithm, a comparable image quality at reduced radiation doses can be reached. In the other hand, the use of the ADMIRE algorithm improved image quality parameters, among others the spatial resolution, for the equal radiation dose. \\ \\
\textit{Key words}: Dual-source CT; ADMIRE; Image texture analysis; Image visual impression; Haralick texture parameters; structural similarity index
\end{minipage}
\end{abstract}

%%%%%%%%%%%%%%%%%%%%%%%%%%%%%%%%%%%%%%%%%%%%%%%%%%%%%%%%%%%%%%%%%%%%%%%%%%%%%%%%%%%%%%%%%%%%%%%%%%%%%%%%
%%%%%%%%%%%%%%%%%%%%%%%%%%%%%%%%%%%%%%%%%%%%%%%%%%%%%%%%%%%%%%%%%%%%%%%%%%%%%%%%%%%%%%%%%%%%%%%%%%%%%%%%
%%%%%%%%%%%%%%%%%%%%%%%%%%%%%%%%%%%%%%%%%%%%%%%%%%%%%%%%%%%%%%%%%%%%%%%%%%%%%%%%%%%%%%%%%%%%%%%%%%%%%%%%
\twocolumn
\section{Introduction}
\label{sec:Introduction}

The use of the X-ray computed tomography (CT) has increased considerably in recent years worldwide which led to the increasing the radiation dose to patients related with the CT\,\cite{ICRU87,kalender_2014}. The total number of CT procedures performed annually in the United States has increased from approximately 3 million in 1980 to about 70 million in 2007\,\cite{Bindman_2009}. Furthermore, the new researches report that 78.7 million CT examinations were performed in the United States in 2015\,\cite{IMV_2015}. Although in the United States only 11-13\% of radiologic examinations have been attributed to the CT, CT is accountable for more than two thirds of the collective effective dose associated with the X-ray imaging\,\cite{Hara_2009}. A similar trend has been observed in Germany: CT examinations are responsible for nearly 9\% of the total X-ray examinations which contributes approximately 65\% of the collective effective dose with the X-ray imaging in Germany\,\cite{BfS_2016}. These reports indicate that a reduction of the radiation dose related to the CT examinations is indispensable. An alternative to reduced the radiation dose in the CT is the use of a lower X-ray tube current, which increases simultaneously the image noise by the use of filtered back projection algorithms (FBP)\,\cite{Patino_2015}. In order to overcome this limitation, CT manufacturers have developed iterative reconstruction (IR) algorithm for conserving image quality by the radiation dose reduction. Several studies demonstrated the performance of the IR algorithms and the radiation dose reduction compared with the FBP technique\,\cite{Patino_2015,Baker_2012,Kalra_2013,vonFalck_2013,McCollough_2015,Solomon_2015}. An iterative algorithm, advanced modeled iterative reconstruction (ADMIRE; Siemens Healthcare, Forchheim, Germany), is becoming clinically available for dual-source CT scanners. In previous studies the effect of ADMIRE on image quality using physical parameters and its dose reduction potential was evaluated comprehensively\,\cite{Solomon_2015,Solomon_2015A,Scholtz_2016,Solomon_2017}. Up to now, to our knowledge, an image quality analysis and a realistic dose reduction potential based on the texture analysis and the visual impression has never been studied. The purpose of the current investigation is to evaluate the impact of ADMIRE on the image texture using the Haralick texture parameters and on the visual impression using the structural similarity index (SSIM) for a dual-source multi-detector CT scanner. Furthermore, the noise distribution and the spatial resolution for diverse radiation dose levels and ADMIRE are discussed.

%%%%%%%%%%%%%%%%%%%%%%%%%%%%%%%%%%%%%%%%%%%%%%%%%%%%%%%%%%%%%%%%%%%%%%%%%%%%%%%%%%%%%%%%%%%%%%%%%%%%%%%%
%%%%%%%%%%%%%%%%%%%%%%%%%%%%%%%%%%%%%%%%%%%%%%%%%%%%%%%%%%%%%%%%%%%%%%%%%%%%%%%%%%%%%%%%%%%%%%%%%%%%%%%%
%%%%%%%%%%%%%%%%%%%%%%%%%%%%%%%%%%%%%%%%%%%%%%%%%%%%%%%%%%%%%%%%%%%%%%%%%%%%%%%%%%%%%%%%%%%%%%%%%%%%%%%%

\section{Materials and Methods}
\label{sec:MaterialsMethods}

\subsection{CT Scanner}
\label{subsec:CT_Scanner}

The clinical scanner used in this investigation was a 192-slice dual-source CT (DSCT) of the third generation (Somatom Force; Siemens Healthcare, Forchheim, Germany). The image acquisition protocols (IAPs) for the measurements are listed in Table \,\ref{tab:IAPs}. In order to obtain a constant radiation dose during the gantry rotation, the radiation dose modulation was disabled. For the image reconstruction an abdomen protocol with a standard kernel (Bf44) and the ADMIRE algorithms with five reconstruction strengths (I-V) were utilized. 
\begin{table} [!ht]
\small
\centering
\caption{IAPs for the presented investigation \label{tab:IAPs}}
\begin{tabularx}{\linewidth}{lll} \hline \hline
X-ray tube peak voltage	&	&	120\,kV				\\ 
X-ray tube current				&	&	20, 50, 100 and 200\,mA		\\
$\mathrm{CTDI_{vol}}$		&	&	0.67, 1.64, 3.31 and 6.65\,mGy	\\
Slice thickness					&	&	1\,mm				\\
Pitch				&	&	1				\\
Exposure time					&	&	0.5\,s				\\
Reconstruction algorithm	&	&	FBP, ADMIRE I-V			\\
Reconstruction kernel				&	&	Bf44				\\ \hline \hline
\end{tabularx}
\end{table}

% ------------------------------------------------------------------------------------------------------

\subsection{Phantom}
\label{subsec:Phantom}

\begin{figure}[!ht]
\centering
\includegraphics[width=0.395\textwidth]{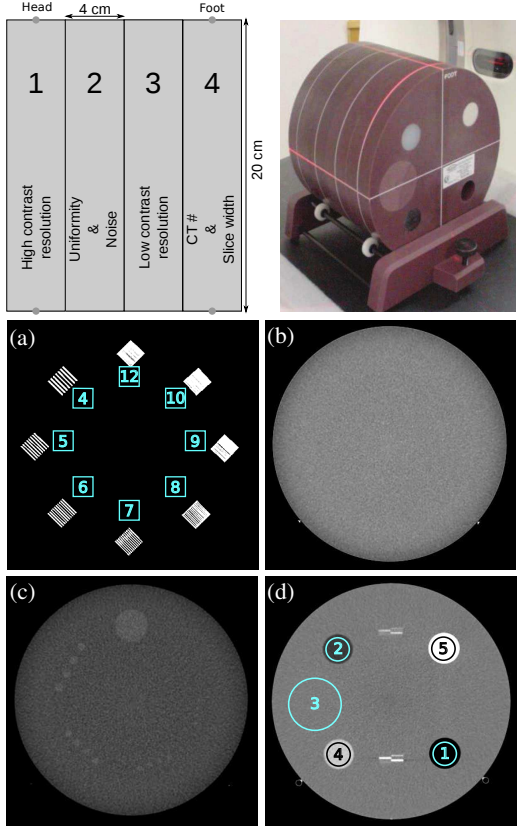}
\caption{ACR phantom. (a) module~1 for spatial resolution. The numbers in square present the count of line pairs per cm ($\mathrm{lp/cm}$). (b) module~2 for noise determination. (c) module~3 for calculation of the Haralick parameters and $\mathrm{SSIM}$. (d) module~4. The numbers indicate different materials, 
\ding{192}:\,Air, 
\ding{193}:\,Polyethylene, 
\ding{194}:\,Water-equivalent material, 
\ding{195}:\,Acrylic, 
\ding{196}:\,Bone. For the purpose of this investigation, the images of the module~4 were not utilized.
 \label{fig:ACR_Phantom_Sketch}}
\end{figure}

An ACR phantom was used to determine the image quality parameters discussed in this work. The ACR phantom consists of a water-equivalent material and contains four modules for the image quality analysis\,\cite{ACR_CT_Phantom_1,ACR_CT_Phantom_2,McCollough_2004}. A sketch, a photograph and CT images of the ACR phantom modules are shown in Figure\,\ref{fig:ACR_Phantom_Sketch}. Images of the ACR phantom were exported via DICOM (Digital Imaging and Communication in Medicine) and analyzed using the software program ImageJ (open-source image analysis software, version 1.50d; \url{https://www.imagej.nih.gov/ij/})

% ------------------------------------------------------------------------------------------------------

\subsection{Haralick texture parameters}
\label{subsec:Haralick_texture_parameters}

The statistical analysis delivers a possibility for the texture analysis of a digital images. This kind of analysis based on not only individual pixel values, but correlations between gray value combinations of pixels. Using gray level matrices, the so-called co-occurrence matrices, different texture parameters introduced by Haralick \textit{et al.} can be calculated\,\cite {Haralick_1973}. A gray level matrix $M_\mathrm {d}(i_1, i_2)$ describes the occurrence rate of the gray values ​​$i_1$ and $i_2$ of two pixels at a distance $\mathrm{d}$ and is defined as
\begin{scriptsize}
\begin{eqnarray*} \label{eq:Co_Occurrence_Matrizen}
M_\mathrm{d}(i_1,i_2) = \left(
\begin{matrix}
p_\mathrm{d}(0,0)	& p_\mathrm{d}(1,0)	& \cdots & p_\mathrm{d}(N_g-1,0) \\ 
p_\mathrm{d}(0,1)	& p_\mathrm{d}(1,1)	& \cdots & p_\mathrm{d}(N_g-1,1) \\  
\vdots 			& \vdots 	  	& \ddots & \vdots 		\\  
p_\mathrm{d}(0,N_g-1)	& p_\mathrm{d}(1,N_g-1)	& \cdots & p_\mathrm{d}(N_g-1,N_g-1)
\end{matrix}
\right) ,
\end{eqnarray*}
\end{scriptsize}
where $p_\mathrm{d}(i, j)$ and $N_g$ characterize the occurrence probability of two gray values ​​$i_1$ and $i_2$ at the distance $\mathrm{d}$ and the possible gray values, respectively. The probabilities $p_\mathrm{d}(i, j)$ are the basic of the various Haralick texture parameters and can be described as follows
\begin{eqnarray}
 0 ~ \leq p_\mathrm{d}(i_1,i_2) ~ \leq 1 ~~~ \forall ~ i_1,\,i_2 ~, \\
 \sum_{i_1=0}^{N_g - 1} \sum_{i_2=0}^{N_g - 1}p_\mathrm{d}(i_1,i_2) = 1 ~.
\end{eqnarray}
% The gray level matrix $M_\mathrm {d}(i_1, i_2)$ depends on the direction choice of the distance $\mathrm{d}$ and is therefore not rotationally invariant. The rotation invariance of the matrix can be ensured by the translation into the polar coordinates (length $ \delta $ and angle $ \theta $) and summing over all possible angles
% \begin{eqnarray}
%  M_\mathrm {d}(i_1,i_2) = \frac{1}{4} \sum_{\theta} M_{\delta,\theta}(i_1,i_2) ~.
% \end{eqnarray}
To characterize the local texture properties, the computation of the gray value matrix is ​​limited to the $ N_g\,\times\,N_g$ image pixels in neighborhood of the considered pixel at a distance $\mathrm {d}$. By the use of the $ N_g\,\times\,N_g$ gray values, 14 texture parameters, the Haralick texture features, are defined, those can adequately describe the properties of the image texture. \\
In order to limit the analysis in this work, two Haralick texture features are just presented, i.e. contrast and entropy. \\
Contrast is a measure of the mean size of gray scale variations in a region. The contrast is defined by
\begin{eqnarray} \label{eq:Haralick_Kontrast}
 C = \sum_{i_1}\sum_{i_2} (i_1-i_2)^2 ~ M_\mathrm {d}(i_1,i_2) ~.
\end{eqnarray}
Image regions with high contrast show a strong gray value changes between pixels. \\
Entropy is the amount of disorder in an image and is given by
\begin{eqnarray} \label{eq:Haralick_Entropie}
 E = \sum_{i_1}\sum_{i_2} \ln(M_\mathrm {d}(i_1,i_2)) ~ M_\mathrm {d}(i_1,i_2) ~.
\end{eqnarray}
The value of entropy reaches a maximum value if all contributions of the gray value matrix are equal. In the case of unequal matrix elements, the value of the entropy is minimal\,\cite{Zayed_2015,Pohle_2004}. \\ \\
For determining of the contrast and entropy, the module~3 of the ACR Phantom were used. This module has a series of cylinders with different diameters and CT numbers close to that of the background, $\mid \Delta \mathrm{CT}\# \mid\, \approx \, 6 \,\mathrm{HU}$.. 

% ------------------------------------------------------------------------------------------------------

\subsection{Structural similarity index}
\label{sec:MM_structural_similarity_index}

% In addition to the module~3, the module~2 with a uniform, tissue-equivalent material was used to calculate SSIM. 

The structural similarity index ($\mathrm{SSIM}$) proposed by Wang \textit{et~al.}\,\cite{Wang_2004} provides a good approximation of perceived image quality which is based on the analysis of the luminance, contrast and structural similarity of two images. In this study $\mathrm{SSIM}$ was used to compare the visibility difference of images with a reference image. $\mathrm{SSIM}$ between two images quantifies their similarity which is denoting by one for identical and smaller than one for non-identical images. \\
In this work $\mathrm{SSIM}$ for two different data sets was determined: First, for each data set measured by a certain tube current, an image reconstructed by the FBP algorithm as the reference image was selected. $\mathrm{SSIM}$ between images reconstructed by the ADMIRE levels and the reference image was calculated ($\mathrm{SSIM_{FBP}}$). For the second analysis, an image measured by a tube current of 900\,mA and reconstructed by FBP BF44 was chosen as a reference image. $\mathrm{SSIM}$ between the reference image and images measured by all tube currents and reconstructed by FBP and ADMIRE levels was determined ($\mathrm{SSIM_{R}}$).\\
Furthermore, $\mathrm{SSIM}$ was used to express the visibly perceivable difference between the images including virtual lesions with different area and the corresponding lesion-free images ($\mathrm{SSIM_L}$). For this purpose, virtual lesions with areas corresponded to circles with diameters of 4, 8, 12, 16 and 20\,mm were inserted to the module~2 with a uniform, tissue-equivalent material. The lesions were added to the background successively and $\mathrm{SSIM_L}$ was calculated by presence of only a lesion. The virtual lesions were inserted to the background images with a gamma-value of 1.03 ($\gamma$ = 1.03). Thereby, the intensity of each pixel ($p$) in the virtual lesions were raised to the power of the gamma-value. For 8-Bit images the new intensity of pixels is given by a function $f(p)$ that is defined as
\begin{eqnarray}
 f(p) = 255 \, \cdot \, (p/255)^{\gamma}.
\end{eqnarray}
In order to calculate $\mathrm{SSIM}$, a tool developed by Kornel was used (\url{https://github.com/kornelski}). This tool computes dissimilarity ($\mathrm{dSSIM}$) between two images. So $\mathrm{SSIM}$ is given by
\begin{eqnarray}
\mathrm{SSIM} = \frac{1}{\mathrm{dSSIM}+1} ~.
\end{eqnarray}
$\mathrm{dSSIM}$ for two identical images is 0, and values greater than 0 denote amount of difference. 

% ------------------------------------------------------------------------------------------------------

\subsection{Image noise determination}
\label{subsec:MM_Image_noise_determination}

By the use of the uniform module~2 of the ACR phantom, the influence of the radiation dose and the ADMIRE levels on the image noise was analyzed. For the determination of the image noise, images of the uniformity module~for all tube currens and the ADMIRE levels were subtracted from a gold-standard image. The gold-standard image was measured by a tube current of 900\,mA and reconstructed using the FBP algorithm. The CT number (HU) distributions of the subtracted images represented the impact of the radiation dose as well as the reconstruction algorithm on the noise.

% ------------------------------------------------------------------------------------------------------

\subsection{High contrast resolution}
\label{sec:MM_High_contrast_resolution}

The modulation transfer function (MTF) is a physical characteristic which describes the high contrast or spatial resolution of an imaging system. A common method to compute the MTF is the Fourier transform calculation of the line-spread function (LSF). The MTF may also be determined with a technique introduced by Droege and Morin\,\cite{Droegen_Morin_1982}. This method is based on the standard deviation measurements of the pixel values within an image of bar patterns. For the MTF determination in this study the technique by Droege and Morin was used. Thereby, ROIs with an area of about 100\,$\mathrm{mm}^2$ within all bar patterns in the module~4 and in the background were placed. Using the CT numbers and standard deviations in the ROIs and the background, MTF for each spatial frequencies (lp/cm) was determined. \\
In addition to MTF measurements, LSF using two very small beads (0.28\,mm each) located in the module~2 was calculated. The full width at half maximum (FWHM) of LSF can be also observed as an indicator for the spatial resolution of the system. In the presented study, MTF as well as FWHM for all radiation dose and ADMIRE levels were assessed.

%%%%%%%%%%%%%%%%%%%%%%%%%%%%%%%%%%%%%%%%%%%%%%%%%%%%%%%%%%%%%%%%%%%%%%%%%%%%%%%%%%%%%%%%%%%%%%%%%%%%%%%%
%%%%%%%%%%%%%%%%%%%%%%%%%%%%%%%%%%%%%%%%%%%%%%%%%%%%%%%%%%%%%%%%%%%%%%%%%%%%%%%%%%%%%%%%%%%%%%%%%%%%%%%%
%%%%%%%%%%%%%%%%%%%%%%%%%%%%%%%%%%%%%%%%%%%%%%%%%%%%%%%%%%%%%%%%%%%%%%%%%%%%%%%%%%%%%%%%%%%%%%%%%%%%%%%%

\section{Results}
\label{sec:Results}

% ------------------------------------------------------------------------------------------------------

\subsection{Haralick texture parameters}
\label{subsec:Res_Haralick_texture_parameters}

\begin{figure}[!ht]
\centering
\includegraphics[width=0.495\textwidth]{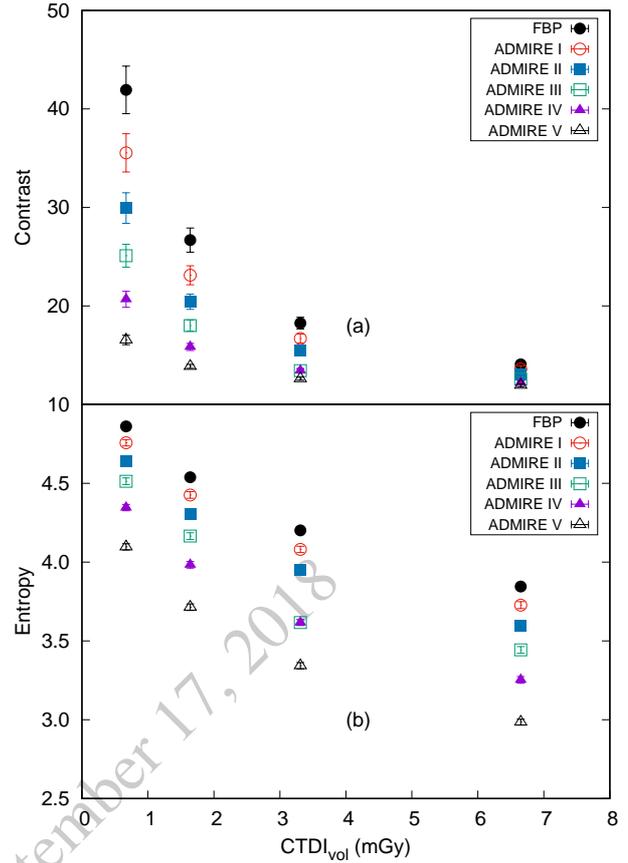}
\caption{(a) Contrast and (b) Entropy calculated for all IAPs. For each data point 35 images were used to calculate the contrast and entropy. \label{fig:Contrast_Entropy}}
\end{figure}

The Haralick texture parameters contrast and entropy calculated for all IAPs are presented in Figures\,\ref{fig:Contrast_Entropy} (a) and (b), respectively. In order to calculate the contrast and entropy in this work, the wighted averages of 35 images for each data point were utilized.
 \\
The figures show a falling trend of the contrast and the entropy with radiation dose. These values also decreased with increasing ADMIRE levels at the same radiation dose. \\
The values of the Haralick contrast, entropy and their uncertainties are listed in Tables\ref{tab:contrast} and \ref{tab:entropy}, respectively, s. Appendix.

% ------------------------------------------------------------------------------------------------------

\subsection{Structural similarity index}
\label{sec:Res_structural_similarity_index}

\begin{figure}[!ht]
\centering
\includegraphics[width=0.495\textwidth]{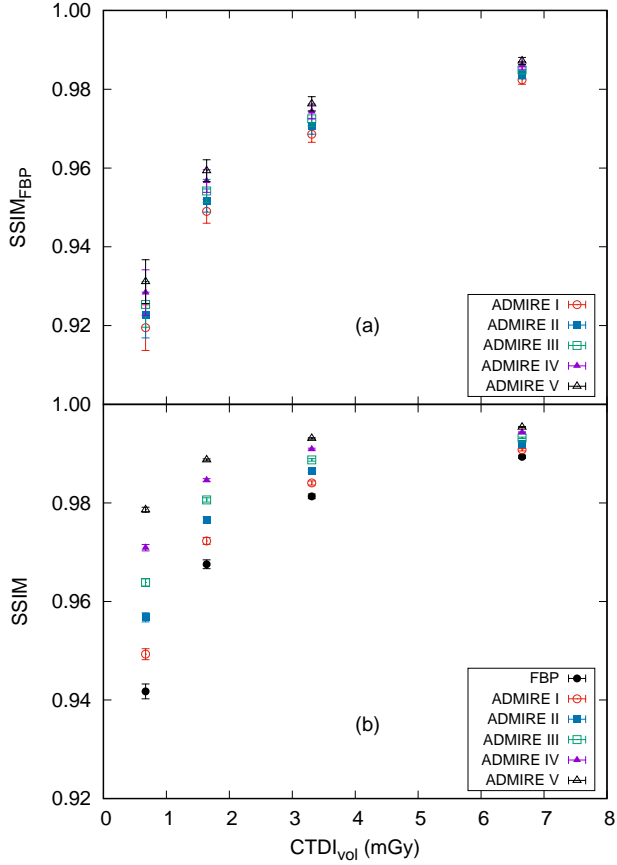}
\caption{(a) $\mathrm{SSIM}$ relative to FBP ($\mathrm{SSIM_{FBP}}$) for measurements with different tube currents and ADMIRE levels. For each data set obtained by identical radiation dose, an image reconstructed by FBP was selected as reference image. (b) $\mathrm{SSIM_{R}}$ as a function of radiation dose  for different reconstruction algorithms. An image reconstructed by FBP Bf44 with a tube current of 900\,mA was chosen as reference image.\label{fig:MS_SSIM}}
\end{figure}

$\mathrm{SSIM_{FBP}}$ for measurements with different radiation dose values is illustrated in Figure\,\ref{fig:MS_SSIM}\,(a).
In Figure\,\ref{fig:MS_SSIM}\,(b) $\mathrm{SSIM_{R}}$ calculated by the use of the reference images and images measured by all tube currents and reconstructed by the FBP and ADMIRE algorithms is presented. As expected, by increasing the radiation dose the visible similarity between images and the reference image has been improved. A similar behavior of $\mathrm{SSIM_{R}}$ with increasing the ADMIRE levels has been founded. \\
$\mathrm{SSIM_{FBP}}$ and $\mathrm{SSIM_{R}}$ were calculated by the weighted mean of 35 pairs of CT images in this work, i.e. 70 images for $\mathrm{dSSIM}$ determination of each data set were used.

% A logarithmic fitting function delivers a good approximation to quantify the dependence of $\mathrm{SSIM}$ on the radiation dose. The fit parameters of a logarithmic fitting function \mbox{(SSIM\,$=\,m \,\cdot\, \mathrm{ln \,\left( CTDI_{vol} \right)}+b$)} and their relative uncertainties are summarized in Table\,\ref{tab:logarithmic_fitting_function}.
% 
% 
% 
% 
% \begin{table} [!ht]
% \centering
% \caption{Fit parameters $m$ and $b$ and the coefficients of determination $R^2$. \label{tab:logarithmic_fitting_function}}
% \begin{tabular}{lccc} \hline \hline
% 				&	$m$		&	$b$		&	$R^2$	\\ \hline
% FBP Bf44	&	0.021$\pm$0.003	&	0.953$\pm$0.003	&	0.964	\\
% 		  \mbox{ADMIRE} 1	&	0.018$\pm$0.003	&	0.960$\pm$0.003	&	0.959	\\
% \mbox{ADMIRE} 2	&	0.015$\pm$0.002	&	0.966$\pm$0.003	&	0.956	\\
% 		  \mbox{ADMIRE} 3	&	0.013$\pm$0.002	&	0.971$\pm$0.002	&	0.952	\\
% \mbox{ADMIRE} 4	&	0.010$\pm$0.002	&	0.977$\pm$0.002	&	0.945	\\
% 		  \mbox{ADMIRE} 5	&	0.007$\pm$0.001	&	0.983$\pm$0.002	&	0.937	\\ \hline \hline
% \end{tabular}
% \end{table}

\begin{figure}[!ht] 
\centering
\includegraphics[width=0.395\textwidth]{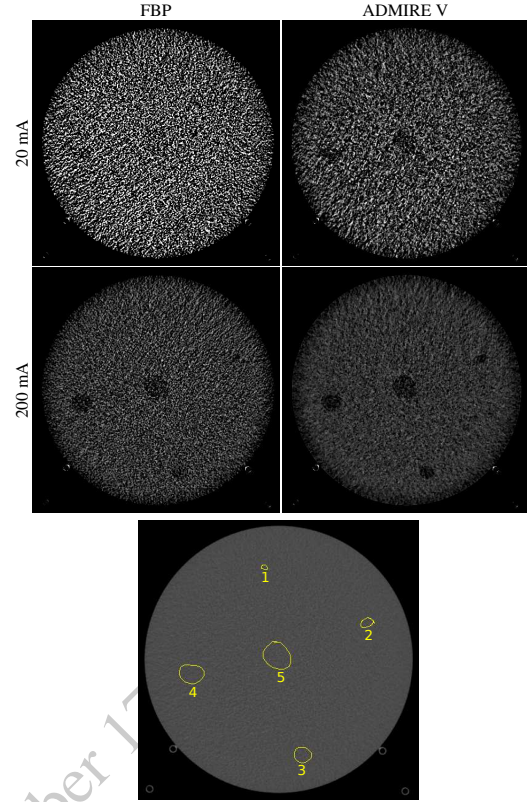}
\caption{Virtual lesion with various areas to the background images measured with tube currents of 20 and 200\,mA and reconstructed by FBP and the ADMIRE level V. The contours of the lesions are shown in the lower figures. \label{fig:lesions_Pics}}
\end{figure}

\begin{figure}[!ht]
\centering
\includegraphics[width=0.495\textwidth]{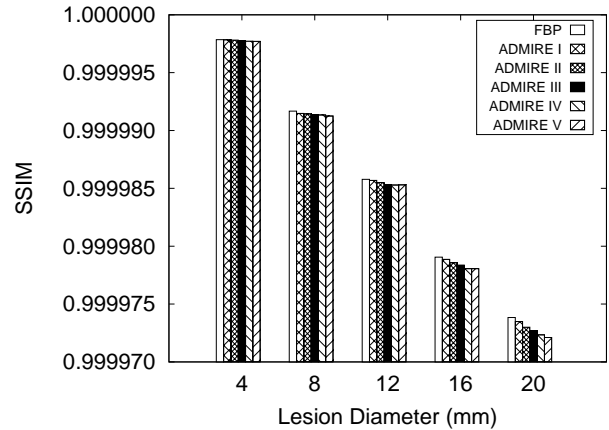}
\caption{$\mathrm{SSIM}$ relative to lesion-free images ($\mathrm{SSIM_L}$ ) as a function of the lesion diameter (area) reconstructed by FBP and ADMIRE for a tube current of 100\,mA. \label{fig:lesions}}
\end{figure}

Finally, Figure\,\ref{fig:lesions_Pics} depicts the background images including virtual lesions of different areas, those correspond to circles with diameters of 4, 8, 12, 16 and 20\,mm. 
The lesion-free images were obtained by tube currents of 20 and 200\,mA. While at the middle of image measured by the tube current of 20\,mA and reconstructed with FBP the lesion with largest area (lesion 5) can be presumed, in the image measured by the same radiation dose and reconstructed with ADMIRE V the lesions 4 and 5 can be seen. In the image measured by the tube current of 200\,mA and reconstructed by FBP the lesions 3, 4 and 5 can be observed. Expect to the lesion 1, other lesions in the image measured by the tube current of 200\,mA and reconstructed by ADMIRE 5 can be easily founded.
$\mathrm{SSIM_L}$ as a function of the lesion diameter and ADMIRE is plotted in Figure\,\ref{fig:lesions} and listed in the Table\,\ref{tab:Lesions}, s. Appendix.
The similarity of image with virtual lesion and lesion-free images were decreased by increasing the lesion areas as well as ADMIRE levels. This result is in accordance with images presented in Figure\,\ref{fig:lesions_Pics}.

% ------------------------------------------------------------------------------------------------------

\subsection{Image noise characteristic}
\label{sec:RES_Image_noise_characteristic}

\begin{figure*}[!ht]
\centering
\includegraphics[width=0.99\textwidth]{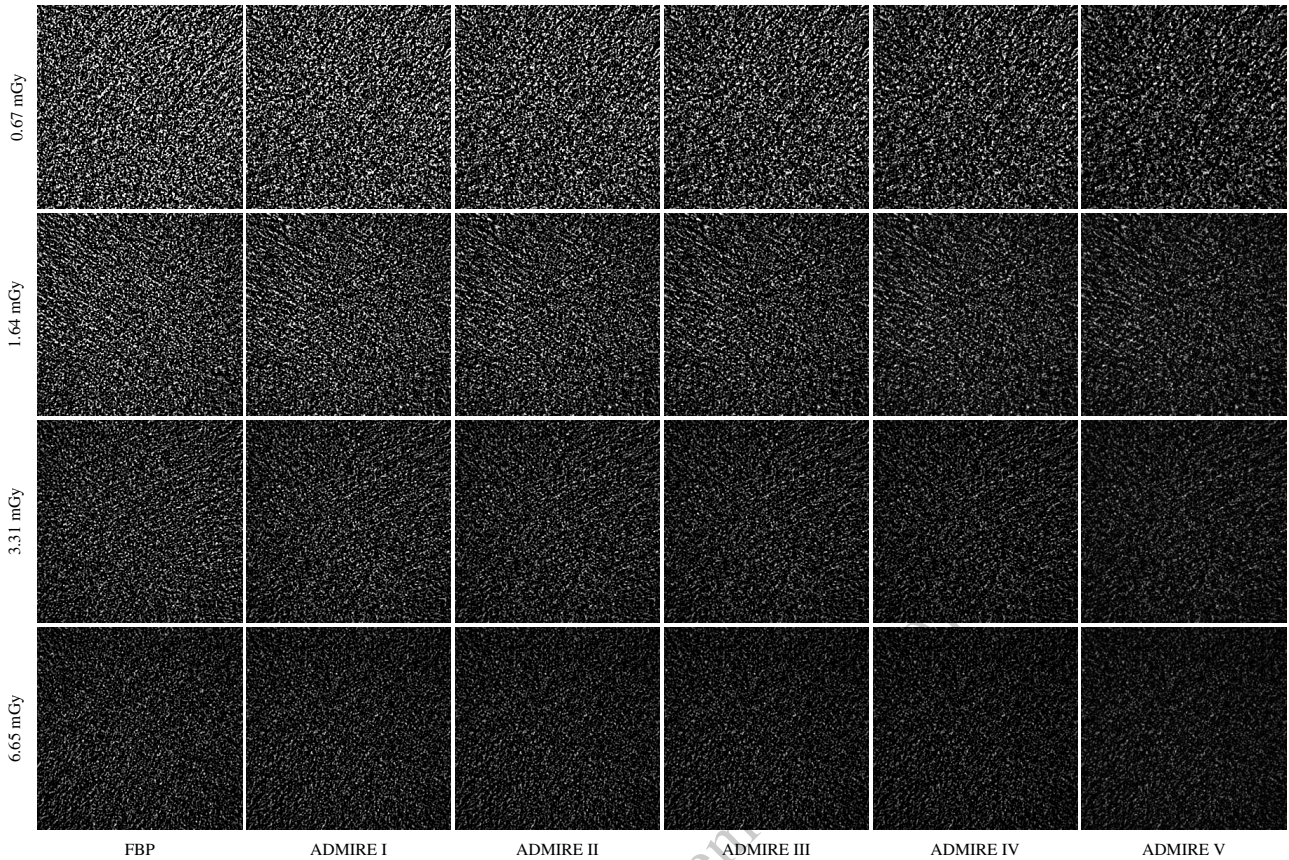}
\caption{Subtracted images from the gold-standard image for the tube currents of 20, 50, 100 and 200\,mA (radiation dose of 0.67, 1.64, 3.31 and 6.65 mGy) reconstructed by FBP and the ADMIRE level I up to V. \label{fig:background}}
\end{figure*}

\begin{figure}[!ht]
\centering
\includegraphics[width=0.495\textwidth]{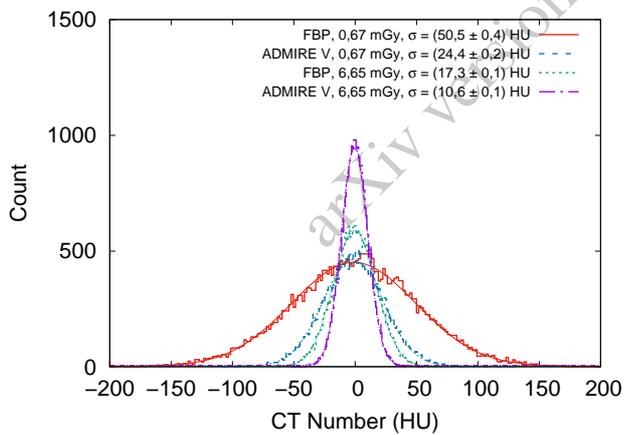}
\caption{Noise distribution for the background images. With increasing the radiation dose and using ADMIRE levels, the HU numbers were shifted to the narrower distribution. The standard deviations of the Gaussian fits are indicated by $\sigma$. \label{fig:HU_density_Results}}
\end{figure}

As mentioned in the section about the image noise calculation, images measured by all data sets were subtracted from a gold-standard image. Figure\,\ref{fig:background} shows subtracted images for the tube currents of 20, 50, 100 and 200\,mA reconstructed by FBP and the ADMIRE level I up to V. This figure shows, that the noise texture was moved to the smoother structure with increasing the ADMIRE level as well the radiation dose. 
Figure\,\ref{fig:HU_density_Results} presented the noise distribution and the corresponding Gaussian fits for the subtracted images presented in Figure\,\ref{fig:background}. With increasing the radiation dose, the HU numbers were relocated to the narrower distribution. Furthermore, the HU numbers distribution were also shifted to the narrower Gaussian by the use of the ADMIRE levels. The center-of-masses for the distributions were almost at the same position by ($-1.2 \pm 0.7$)\,HU. 

% ------------------------------------------------------------------------------------------------------

\subsection{High contrast resolution}
\label{sec:RES_High_contrast_resolution}

\begin{figure}[!ht]
\centering
\includegraphics[width=0.495\textwidth]{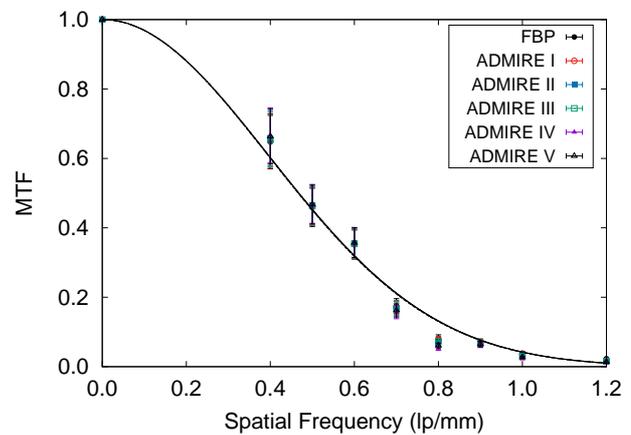}
\caption{MTF measured by a tube current of 100\,mA. An appreciable variation of MTF was not observed.\label{fig:MTF}}
\end{figure}

\begin{figure}[!ht]
\centering 
\includegraphics[width=0.495\textwidth]{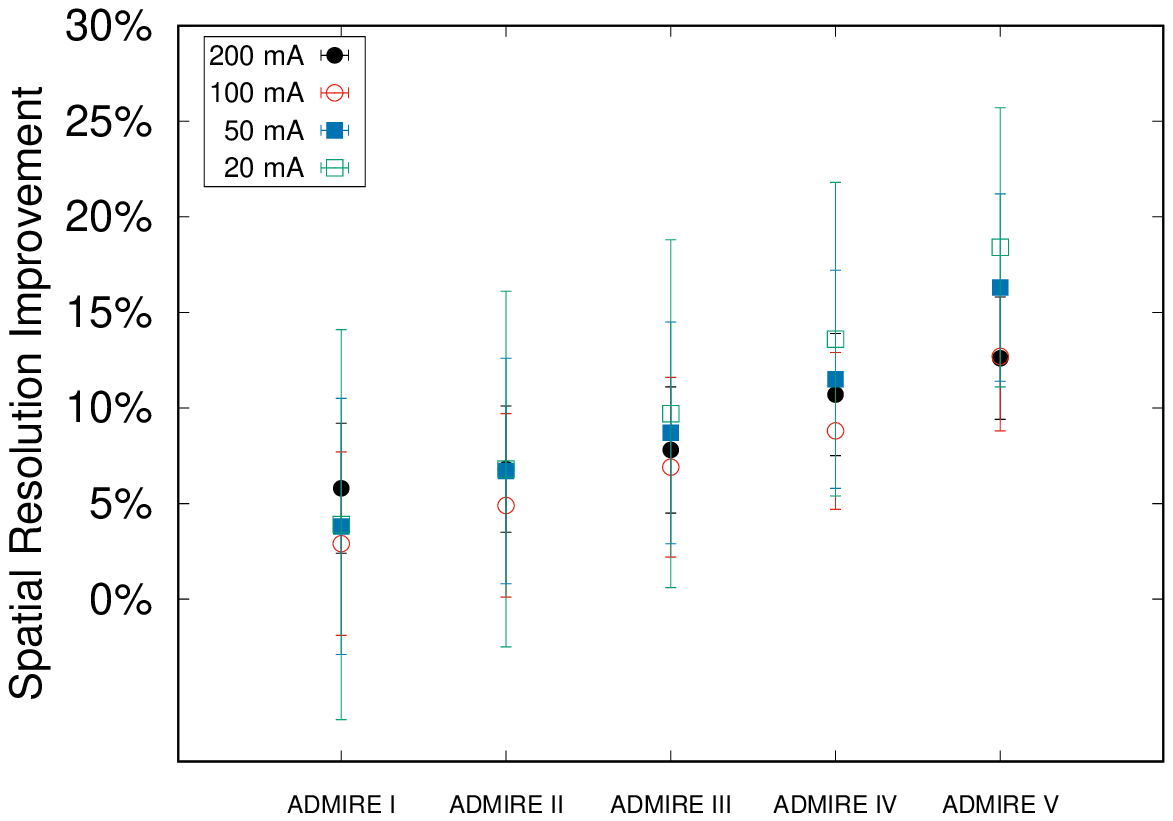}
\caption{Spatial resolution improvement of ADMIRE levels in comparison to FBP for tube currents of 20, 50, 100 and 200\,mA. Using ADMIRE, the spatial resolution has been improved.\label{fig:FWHM_improvement}}
\end{figure}

The MTF values and the correlated fits versus the spatial resolution measured by FBP and the ADMIRE levels for a tube current of 100\,mA are depicted in Figure\,\ref{fig:MTF}. Subsequently, using the beads the LSFs were determined which were fitted by the Gaussian functions. FWHM of the Gaussian fitting functions for all data sets are summarized in Table\,\ref{tab:FWHM}, s. Appendix. Figure\,\ref{fig:FWHM_improvement} shows the spatial resolution improvement of the ADMIRE levels in comparison with FBP for tube currents of 20, 50, 100 and 200\,mA. 
While the spatial resolution variation as a function of the radiation dose could be neglected, it showed an improvement up to 18\,\% by the use of ADMIRE levels.

%%%%%%%%%%%%%%%%%%%%%%%%%%%%%%%%%%%%%%%%%%%%%%%%%%%%%%%%%%%%%%%%%%%%%%%%%%%%%%%%%%%%%%%%%%%%%%%%%%%%%%%%
%%%%%%%%%%%%%%%%%%%%%%%%%%%%%%%%%%%%%%%%%%%%%%%%%%%%%%%%%%%%%%%%%%%%%%%%%%%%%%%%%%%%%%%%%%%%%%%%%%%%%%%%
%%%%%%%%%%%%%%%%%%%%%%%%%%%%%%%%%%%%%%%%%%%%%%%%%%%%%%%%%%%%%%%%%%%%%%%%%%%%%%%%%%%%%%%%%%%%%%%%%%%%%%%%

\section{Discussion}
\label{sec:Discussion}

The presented work showed the texture analysis and an evaluating the visual impression of images provided by a dual-source CT scanner (DSCT) and reconstructed by a model-based iterative reconstruction (ADMIRE) algorithm. The image texture was analyzed using two Haralick parameters, while the image visual impression was assessed in terms of the structural similarity index (SSIM). Furthermore, the noise distributions as well as the spatial resolutions as  functions of the ADMIRE and radiation dose levels were determined. \\ \\
The model-based ADMIRE algorithm has been considerably improved relative to the last-generation IR by the same company, SAFIRE (sinogram-affirmed iterative reconstruction; Siemens Healthcare, Forchheim, Germany)\,\cite{Solomon_2015A}, which is only installed in the second generation of DSCT scanners. SAFIRE was recently evaluated in different studies and its potential for radiation dose reduction was discussed, e.g. by Baker~\textit{et~al.}\,\cite{Baker_2012}, Kalra~\textit{et~al.}\,\cite{Kalra_2013}, von Falck~\textit{et~al.}\,\cite{vonFalck_2013} and McCollough~\textit{et~al.}\,\cite{McCollough_2015}. In an investigation, Solomon~\textit{et~al.} showed the increase of the detectability performance for the SAFIRE algorithm compared with FBP\,\cite{Solomon_2017}. \\
ADMIRE was also discussed in various publications. For example, Solomon~\textit{et~al.}\,\cite{Solomon_2015} demonstrated in a comprehensive study the assessment the image quality in terms of the noise power spectrum, task transfer function and detectability performance as a function of IAPs and different reconstruction algorithms, FPB and ADMIRE strengths III to IV. In another phantom study of the same research group\,\cite{Solomon_2015A}, the performance of ADMIRE for the low-contrast detectability and the potential for radiation dose reduction were evaluated. On the basic of two reading sessions, the data sets were assessed by readers and the dose reduction potential was estimated to be 41\,\% in average by the use of the ADMIRE strengths III to IV in comparison to FBP. Scholtz \textit{et~al.}\,\cite{Scholtz_2016} reported a reduction of the noise and an improvement of CNR and SNR by the use of ADMIRE. Thereby, 116 patients were selected who underwent the CT examination of the neck. \\
Although, in the mentioned studies the characterization of the ADMIRE algorithm based on the image quality parameters, such as noise, CNR and SNR, MTF, low-contrast detectability, or on the reader studies were showed, the image visual impression was not discussed. \\ \\
Up to now, to our knowledge, no study has assessed the impact of the ADMIRE algorithm on the image texture and the image visual impression. Therefore, the aim of the presented study included the investigating the ADMIRE influence on the image texture and the image visual impression of images provided by a dual-source multi-detector CT scanner. As for the texture analysis, two Haralick texture parameters, contrast and entropy, were only utilized, the image visual impression was characterized by $\mathrm{SSIM}$. $\mathrm{SSIM}$ delivers a more realistic assessment of the dose reduction potential of iterative reconstruction algorithms than quality metrics only based on physical measurements of spatial and contrast resolution. $\mathrm{SSIM}$ can be also viewed as an index to evaluate the low-contrast detectability\,\cite{vonFalck_2012,vonFalck_2013}. Furthermore, the determination of $\mathrm{SSIM}$ as a detectability indicator can replace the reader studies, which is in general sophisticated. \\
$\mathrm{SSIM}$ (SSIM) showed an enhancement with increasing the ADMIRE level which is more obvious for ADMIRE IV and V. An increasing from 18\,\% (ADMIRE I, 20\,mA) up to 23\,\% (ADMIRE V, 200\,mA) with the radiation dose values could be observed. \\
The calculated $\mathrm{SSIM}$ for images and a reference image exhibited that the same $\mathrm{SSIM}$ for the ADMIRE III achieved at about 50\,\% radiation dose when compared with FBP obtained at 100\,\%.
Furthermore, $\mathrm{SSIM}$ showed a decreasing trend with the ADMIRE levels for all lesion diameters, i.e. the lesions in the images reconstructed by the ADMIRE IV were most visible. \\
Although the potential of the radiation dose reduction and improvement of the low-contrast detectability of the basis of $\mathrm{SSIM}$ are in accordance with the results obtained by the reader studies mentioned in previous studies, $\mathrm{SSIM}$ provides an easier and more efficient method to assess the image quality. The texture analysis provides similar results. Using ADMIRE levels, the variation of the gray values (contrast) and the disorder (entropy) in a image were decreasing. While the mean CT numbers of background figured out the weakly dependence of reconstruction algorithm and can be assumes as a constant, their standard derivations (noise) were decrease with the ADMIRE levels. This result was also observed in the previous studies\,\cite{Solomon_2015,Scholtz_2016}. The HU density histograms of the noise showed a narrower distribution by increasing the radiation dose. In addition, the use of the ADMIRE levels led to decrease of the image noise and narrower distribution and the image noise texture shifted to the smoother structures with increase of the ADMIRE levels which could be observed for all radiation dose. A significant difference of MTF for all IAPs and reconstruction algorithms was not observed. On the other hand, by increasing the ADMIRE levels the FWHM values of the LSF showed a decreasing trend. This result indicates an improvement of the high contrast resolution with ADMIRE levels. Considering the standard derivation, the spatial resolution was almost independent on the radiation dose level for all reconstruction algorithms. The FHWM calculated by ADMIRE V showed an improvement of 13(3)\,\% and 18(7)\,\% compared with FBP for tube currents of 20\,mA and 200\,mA, respectively. The spatial resolution enhancement by ADMIRE levels is in accordance to the results given by Solomon~\textit{et~al.}\,\cite{Solomon_2015}. Indeed, von Falck~\textit{et~al.}\,\cite{vonFalck_2013} showed that there was no statistically significant difference in spatial resolution by means of SAFIRE algorithm. \\ \\
Our study has a number of limitations: First, we used a basic cylinder-shaped phantom with a radius of 20 \,cm which is only an approximation for a patient body. Second, the radiation dose was equated to the volumetric CT dose index ($\mathrm{CTDI_{vol}}$) which is only the radiation dose output of the CT scanners and proportional the patient dose. A dose measurement using TLD or an appropriate ionization chamber might have been beneficial. However, this is time consuming and would in our opinion not alter the results. \\ \\
In conclusion, this work presented the influence of the ADMIRE algorithm on the visual impression and image noise. The structural similarity evaluation showed that the low contrast detectability using ADMIRE at 50\,\% radiation dose was almost equal to that by the use of FBP at 100\,\% radiation dose. In addition, the results indicated that the spatial resolution improved by the use of ADMIRE algorithm.

%%%%%%%%%%%%%%%%%%%%%%%%%%%%%%%%%%%%%%%%%%%%%%%%%%%%%%%%%%%%%%%%%%%%%%%%%%%%%%%%%%%%%%%%%%%%%%%%%%%%%%%%
%%%%%%%%%%%%%%%%%%%%%%%%%%%%%%%%%%%%%%%%%%%%%%%%%%%%%%%%%%%%%%%%%%%%%%%%%%%%%%%%%%%%%%%%%%%%%%%%%%%%%%%%
%%%%%%%%%%%%%%%%%%%%%%%%%%%%%%%%%%%%%%%%%%%%%%%%%%%%%%%%%%%%%%%%%%%%%%%%%%%%%%%%%%%%%%%%%%%%%%%%%%%%%%%%

% \bibliographystyle{unsrt}
% \bibliography{literatur_SiemensForce_ADMIRE_Effect}

\begin{thebibliography}{10}

\bibitem{ICRU87}
{ICRU REPORT} {N}o. 87: {R}adiation {D}ose and {I}mage-{Q}uality {A}ssessment
  in {C}omputed {T}omography.
\newblock {\em The International Commission on Radiation Units and
  Measurements}, 12(1), 2012.

\bibitem{kalender_2014}
Kalender WA.
\newblock Dose in x-ray computed tomography.
\newblock {\em Physics in Medicine and Biology}, 59(3):R129, 2014.

\bibitem{Bindman_2009}
Smith-Bindman R, Lipson J, Marcus R, et~al.
\newblock Radiation {D}ose {A}ssociated with {C}ommon {C}omputed {T}omography
  {E}xaminations and the {A}ssociated {L}ifetime {A}ttributable {R}isk of
  {C}ancer.
\newblock {\em Archives of internal medicine}, 169(22):2078--2086, 2009.

\bibitem{IMV_2015}
{IMV} {M}edical {I}nformation {D}ivision, {CT} {M}arket {O}utlook {R}eport
  \url{http://www.imvinfo.com/index.aspx?sec=ct&sub=dis&itemid=200081}, 2015.

\bibitem{Hara_2009}
Hara AL, Paden RG, Silva AC, et~al.
\newblock Iterative {R}econstruction {T}echnique for {R}educing {B}ody
  {R}adiation {D}ose at {CT}: {F}easibility {S}tudy.
\newblock {\em American Journal of Roentgenology}, 193(3):764--771, 2009.

\bibitem{BfS_2016}
Umweltradioaktivität und {S}trahlenbelastung, {J}ahresbericht 2014,
  korrigierte {F}assung vom 18. {O}ktober 2016, 2016.

\bibitem{Patino_2015}
Patino M, Fuentes JM, Singh St, et~al.
\newblock Iterative {R}econstruction {T}echniques in {A}bdominopelvic {CT}:
  {T}echnical {C}oncepts and {C}linical {I}mplementation.
\newblock {\em American Journal of Roentgenology}, 205(1):W19--W31, 2015.

\bibitem{Baker_2012}
Baker ME, Dong F, Primak A, et~al.
\newblock Contrast-to-{N}oise {R}atio and {L}ow-{C}ontrast {O}bject
  {R}esolution on {F}ull- and {L}ow-{D}ose {MDCT}: {SAFIRE} {V}ersus {F}iltered
  {B}ack {P}rojection in a {L}ow-{C}ontrast {O}bject {P}hantom and in the
  {L}iver.
\newblock {\em American Journal of Roentgenology}, 199(1):8--18, 2012.

\bibitem{Kalra_2013}
Kalra MK, Woisetschläger M, Dahlström N, et~al.
\newblock Sinogram-{A}ffirmed {I}terative {R}econstruction of {L}ow-{D}ose
  {C}hest {CT}: {E}ffect on {I}mage {Q}uality and {R}adiation {D}ose.
\newblock {\em American Journal of Roentgenology}, 201(2):W235--W244, 2013.

\bibitem{vonFalck_2013}
von Falck~C, Bratanova V, Rodt T, et~al.
\newblock Influence of {S}inogram {A}ffirmed {I}terative {R}econstruction of
  {CT} {D}ata on {I}mage {N}oise {C}haracteristics and {L}ow-{C}ontrast
  {D}etectability: {A}n {O}bjective {A}pproach.
\newblock {\em PLoS ONE}, 8(2):1--10, 2013.

\bibitem{McCollough_2015}
McCollough CH, Yu~L, Kofler JM, et~al.
\newblock Degradation of {CT} {L}ow-{C}ontrast {S}patial {R}esolution {D}ue to
  the {U}se of {I}terative {R}econstruction and {R}educed {D}ose {L}evels.
\newblock {\em Radiology}, 276(2):499--506, 2015.

\bibitem{Solomon_2015}
Solomon J, Joshua Wand, and Samei E.
\newblock Characteristic image quality of a third generation dual-source {MDCT}
  scanner: {N}oise, resolution, and detectability.
\newblock {\em Medical Physics}, 42(8):4941--4953, 2015.

\bibitem{Solomon_2015A}
Solomon J, Mileto A, Ramirez-Giraldo JC, et~al.
\newblock Diagnostic {P}erformance of an {A}dvanced {M}odeled {I}terative
  {R}econstruction {A}lgorithm for {L}ow-{C}ontrast {D}etectability with a
  {T}hird-{G}eneration {D}ual-{S}ource {M}ultidetector {CT} {S}canner:
  {P}otential for {R}adiation {D}ose {R}eduction in a {M}ultireader {S}tudy.
\newblock {\em Radiology}, 275(3):735--745, 2015.

\bibitem{Scholtz_2016}
Scholtz JE, Wichmann JL, Hüsers K, et~al.
\newblock Third-generation dual-source {CT} of the neck using automated tube
  voltage adaptation in combination with advanced modeled iterative
  reconstruction: evaluation of image quality and radiation dose.
\newblock {\em European Radiology}, 26(8):2623--2631, 2016.

\bibitem{Solomon_2017}
Solomon J, Marin Dand~Choudhury KR, et~al.
\newblock Effect of {R}adiation {D}ose {R}eduction and {R}econstruction
  {A}lgorithm on {I}mage {N}oise, {C}ontrast, {R}esolution, and {D}etectability
  of {S}ubtle {H}ypoattenuating {L}iver {L}esions at {M}ultidetector {CT}:
  {F}iltered {B}ack {P}rojection versus a {C}ommercial {M}odel- based
  {I}terative {R}econstruction {A}lgorithm.
\newblock {\em Radiology}, 2017.

\bibitem{ACR_CT_Phantom_1}
{CT} {A}ccreditation {P}hantom {I}nstructions,
  \url{http://www.acr.org/~/media/ACR/Documents/Accreditation/CT/PhantomTestingInstruction.pdf}.

\bibitem{ACR_CT_Phantom_2}
{CT} {A}ccreditation {P}rogram: {I}mage {Q}uality and {D}ose {M}easurements,
  \url{https://www.aapm.org/meetings/03AM/pdf/9785-27333.pdf}.

\bibitem{McCollough_2004}
McCollough CH, Bruesewitz MR, McNitt-Gray MF, et~al.
\newblock The phantom portion of the american college of radiology ({ACR})
  {C}omputed {T}omography ({CT}) accreditation program: Practical tips,
  artifact examples, and pitfalls to avoid.
\newblock {\em Medical Physics}, 31(9), 2004.

\bibitem{Haralick_1973}
RM~Haralick, K~Shanmugam, and I~Dinstein.
\newblock Textural features for image classification.
\newblock {\em IEEE TRANSACTIONS ON SYSTEMS, MAN, AND CYBERNETICS},
  SMC-3(6):610--621, 1973.

\bibitem{Zayed_2015}
N~Zayed and HA~Elnemr.
\newblock Statistical {A}nalysis of h{}aralick {T}exture {F}eatures to
  {D}iscriminate {L}ung {A}bnormalities.
\newblock {\em International Journal of Biomedical Imaging}, ID267807, 2015.

\bibitem{Pohle_2004}
R~Pohle.
\newblock Computerunterstützte b{}ildanalyse zur a{}uswertung medizinischer
  {B}ilddaten. {H}abilitationsschrift.
\newblock 2004.

\bibitem{Wang_2004}
Wang Z, Bovik AC, Sheikh HR, et~al.
\newblock Image quality assessment: from error visibility to structural
  similarity.
\newblock {\em IEEE Trans Image Process}, 13(4):600--12, 2004.

\bibitem{Droegen_Morin_1982}
Droege RT and Morin RL.
\newblock A practical method to measure the {MTF} of {CT} scanners.
\newblock {\em Medical Physics}, 9(5), 1982.

\bibitem{vonFalck_2012}
von Falck~C, Rodt T, Waldeck S, et~al.
\newblock A systematic approach towards the objective evaluation of
  low-contrast performance in {MDCT}: Combination of a full-reference image
  fidelity metric and a software phantom.
\newblock {\em European Journal of Radiology}, 81(11):3166 -- 3171, 2012.

\end{thebibliography}

%%%%%%%%%%%%%%%%%%%%%%%%%%%%%%%%%%%%%%%%%%%%%%%%%%%%%%%%%%%%%%%%%%%%%%%%%%%%%%%%%%%%%%%%%%%%%%%%%%%%%%%%
%%%%%%%%%%%%%%%%%%%%%%%%%%%%%%%%%%%%%%%%%%%%%%%%%%%%%%%%%%%%%%%%%%%%%%%%%%%%%%%%%%%%%%%%%%%%%%%%%%%%%%%%
%%%%%%%%%%%%%%%%%%%%%%%%%%%%%%%%%%%%%%%%%%%%%%%%%%%%%%%%%%%%%%%%%%%%%%%%%%%%%%%%%%%%%%%%%%%%%%%%%%%%%%%%

% \appendix
\section*{Appendix}

\begin{landscape}
% \onecolumn

\begin{table}
% \footnotesize

% ------------------------------------------------------------------------------------------------------

% \centering
\caption{Haralick texture parameter contrast for all dose levels reconstruction algorithms. \label{tab:contrast}}
\begin{tabular}{ccccccc} \hline \hline
$\mathrm{CTDI_{vol}}$	&	FBP		&	ADMIRE I	&	ADMIRE II	&	ADMIRE III	&	ADMIRE IV	&	ADMIRE V		\\ \hline
0.67	&	41.94 $\pm$ 2.41&	35.54 $\pm$ 1.94&	29.94 $\pm$ 1.55&	25.10 $\pm$ 1.16&	20.67 $\pm$ 0.81&	16.54 $\pm$ 0.48	\\
		  1.64	&	26.68 $\pm$ 1.22&	23.11 $\pm$ 0.97&	20.43 $\pm$ 0.77&	18.00 $\pm$ 0.56&	15.84 $\pm$ 0.37&	13.86 $\pm$ 0.20	\\
3.31	&	18.25 $\pm$ 0.60&	16.69 $\pm$ 0.46&	15.43 $\pm$ 0.37&	13.42 $\pm$ 0.21&	13.42 $\pm$ 0.21&	12.60 $\pm$ 0.16	\\
		  6.65	&	14.07 $\pm$ 0.38&	13.59 $\pm$ 0.34&	13.03 $\pm$ 0.28&	12.56 $\pm$ 0.22&	12.20 $\pm$ 0.18&	11.94 $\pm$ 0.15	\\ \hline \hline

\end{tabular}

% ------------------------------------------------------------------------------------------------------

% \centering
\caption{Haralick texture parameter entropy for all dose levels reconstruction algorithms. \label{tab:entropy}}
\begin{tabular}{ccccccc} \hline \hline
$\mathrm{CTDI_{vol}}$	&	FBP		&	ADMIRE I	&	ADMIRE II	&	ADMIRE III	&	ADMIRE IV	&	ADMIRE V		\\ \hline
0.67	&	4.86 $\pm$ 0.02	&	4.76 $\pm$ 0.02	&	4.64 $\pm$ 0.02	&	4.51 $\pm$ 0.02	&	4.35 $\pm$ 0.02	&	4.10 $\pm$ 0.02		\\
		  1.64	&	4.54 $\pm$ 0.02	&	4.43 $\pm$ 0.02	&	4.31 $\pm$ 0.02	&	4.16 $\pm$ 0.02	&	3.98 $\pm$ 0.02	&	3.71 $\pm$ 0.02		\\
3.31	&	4.20 $\pm$ 0.02	&	4.08 $\pm$ 0.02	&	3.95 $\pm$ 0.02	&	3.62 $\pm$ 0.02	&	3.62 $\pm$ 0.02	&	3.34 $\pm$ 0.02		\\
		  6.65	&	3.85 $\pm$ 0.02	&	3.73 $\pm$ 0.02	&	3.59 $\pm$ 0.02	&	3.44 $\pm$ 0.02	&	3.25 $\pm$ 0.02	&	2.99 $\pm$ 0.02		\\ \hline \hline

\end{tabular}

% ------------------------------------------------------------------------------------------------------

% \centering
\caption{$\mathrm{SSIM_L}$ for all reconstruction algorithms measured by a X-ray tube of 100~mA ($\mathrm{CTDI_{vol}}=$3.31~mGy). \label{tab:Lesions}}
\begin{tabular}{ccccccc} \hline \hline
			 &	FBP		 &	ADMIRE I	 &	ADMIRE II	 &	ADMIRE III	 &	ADMIRE IV	 &	ADMIRE V		\\ \hline
Lesion1&0.9999979$\pm$0.0000001&0.9999978$\pm$0.0000001&0.9999978$\pm$0.0000001&0.9999978$\pm$0.0000001&0.9999977$\pm$0.0000001&0.9999977$\pm$0.0000001	\\
		  Lesion2&0.9999917$\pm$0.0000002&0.9999915$\pm$0.0000002&0.9999914$\pm$0.0000002&0.9999913$\pm$0.0000003&0.9999913$\pm$0.0000002&0.9999913$\pm$0.0000002	\\
Lesion3&0.9999858$\pm$0.0000003&0.9999857$\pm$0.0000002&0.9999855$\pm$0.0000002&0.9999853$\pm$0.0000002&0.9999853$\pm$0.0000002&0.9999853$\pm$0.0000002	\\
		  Lesion4&0.9999791$\pm$0.0000003&0.9999789$\pm$0.0000003&0.9999786$\pm$0.0000002&0.9999784$\pm$0.0000002&0.9999781$\pm$0.0000002&0.9999781$\pm$0.0000003	\\
Lesion5&0.9999738$\pm$0.0000003&0.9999735$\pm$0.0000003&0.9999730$\pm$0.0000003&0.9999727$\pm$0.0000004&0.9999723$\pm$0.0000003&0.9999721$\pm$0.0000003	\\ \hline \hline
\end{tabular}

% ------------------------------------------------------------------------------------------------------

% \centering
\caption{FWHM calculated for different radiation dose and reconstruction algorithms. The Numbers are given in units of mm. \label{tab:FWHM}}
\begin{tabular}{ccccccc} \hline \hline
$\mathrm{CTDI_{vol}}$	&	FBP		&	ADMIRE I	&	ADMIRE II	&	ADMIRE III	&	ADMIRE IV	&	ADMIRE V		\\ \hline
0.67	&	1.03 $\pm$ 0.08	&	0.99 $\pm$ 0.07	&	0.96 $\pm$ 0.06	&	0.93 $\pm$ 0.06	&	0.89 $\pm$ 0.05	&	0.84 $\pm$ 0.04		\\
		  1.64	&	1.04 $\pm$ 0.05	&	1.00 $\pm$ 0.05	&	0.97 $\pm$ 0.04	&	0.95 $\pm$ 0.04	&	0.92 $\pm$ 0.04	&	0.87 $\pm$ 0.03		\\
3.31	&	1.02 $\pm$ 0.04	&	0.99 $\pm$ 0.03	&	0.97 $\pm$ 0.03	&	0.93 $\pm$ 0.02	&	0.93 $\pm$ 0.02	&	0.89 $\pm$ 0.02		\\
		  6.65	&	1.03 $\pm$ 0.03	&	0.97 $\pm$ 0.02	&	0.96 $\pm$ 0.02	&	0.95 $\pm$ 0.02	&	0.92 $\pm$ 0.02	&	0.90 $\pm$ 0.02		\\ \hline \hline
\end{tabular}
\end{table}
\end{landscape}

% ------------------------------------------------------------------------------------------------------
% ------------------------------------------------------------------------------------------------------
% ------------------------------------------------------------------------------------------------------

\end{document}